\documentclass[prc,floatfix,groupedaddress,nofootinbib,showpacs,preprintnumbers,
amsmath,amssymb,amsfonts,superscriptaddress,widetable] {revtex4}
\usepackage{bm}
\usepackage{mathrsfs}
\usepackage{amssymb}
\usepackage{amsmath}
\usepackage{mathtools}
\usepackage{graphicx}
\usepackage{array}
\usepackage{color}
\usepackage{relsize}
\usepackage{adjustbox}
\def\yR{y_{\raisebox{-0.75pt}{\tiny {\rm R}}}}
\def\rhoz{\rho_{\raisebox{-0.75pt}{\tiny 0}}}
\def\epsz{\varepsilon_{\raisebox{-0.75pt}{\tiny 0}}}
\def\Edens{\mathlarger{\mathlarger{\varepsilon\hspace{0pt}}}}

\begin{document}
\title{Impact of the neutron star crust on the tidal polarizability}
\author{J. Piekarewicz}\email{jpiekarewicz@fsu.edu}
\affiliation{Department of Physics, Florida State University,
               Tallahassee, FL 32306, USA}
\author{F.J. Fattoyev}\email{ffattoyev01@manhattan.edu}
\affiliation{Department of Physics, Manhattan College,
                Riverdale, NY 10471, USA}

\date{\today}



\begin{abstract}
\medskip
\begin{adjustbox}{minipage=0.75\textwidth}
\begin{description}
 \item[Background:]
  The first direct detection of a binary neutron star merger (GW170817) by the LIGO-Virgo
  scientific collaboration has opened the brand new era of multimessenger astronomy. This
  historic detection has been instrumental in providing initial constraints on the tidal
  polarizability (or deformability) of neutron stars. In turn, the tidal polarizability---an
  observable highly sensitive to stellar compactness---has been used to impose limits on
  stellar radii and ultimately on the underlying equation of state (EOS).
\item[Purpose:]
  Besides its strong dependence on the stellar compactness, the tidal polarizability is
  also sensitive to the second tidal Love number $k_{2}$. It is the main purpose of this
  work to perform a detailed study of  $k_{2}$ which, for a given compactness parameter,
  encodes the entire sensitivity of the tidal polarizability to the underlying equation of state.
  In particular, we examine the important role that the crustal component of the EOS plays
  in the determination of $k_{2}$.

\item[Methods:]
  A set of realistic models of the equation of state that yield an accurate description of
  the properties of finite nuclei, support neutron stars of two solar masses, and provide
  a Lorentz covariant extrapolation to dense matter are used to solve both the
  Tolman-Oppenheimer-Volkoff and the differential equation for the induced quadrupole
  gravitational field from which $k_{2}$ is extracted.

\item[Results:]
  Given that the tidal polarizability scales as the fifth power of the compactness parameter,
  a universal relation exists among the tidal polarizability and the compactness parameter
  that is highly insensitive to the underlying equation of state. Thus, besides an extraction
  of the tidal polarizabilities, a measurement of the individual stellar masses is also required
  to impact the mass-radius relation. However, we observe a strong sensitivity of the second
  Love number to the underlying equation of state---particularly to the contribution from the
  inner crust.

\item[Conclusions:]
  Although by itself the tidal polarizability can not contribute to the determination of the
  mass-radius relation, future detections of binary neutron star mergers by the LIGO-Virgo
  collaboration during the third observing run and beyond are poised to provide significant
  constraints on both the tidal polarizabilities and masses of the individual stars, and thus
  ultimately on the mass-radius relation. Yet, subleading corrections to the tidal polarizability
  are encoded in the second Love number $k_{2}$ which displays a large sensitivity to the
  entire---crust-plus-core---equation of state.
\end{description}

\end{adjustbox}
\end{abstract}
\smallskip

\maketitle

\section{Introduction}
\label{sec:introduction}

Nuclear physics plays a predominant role in explaining the behavior of neutron stars. Neutron stars are
fascinating objects that are born from the collapse of massive stars and reach central densities that may
exceed those found in atomic nuclei by up to an order of magnitude. In contrast to the common perception
of neutron stars as  ``dense stars which look like one giant nucleus"\,\cite{Landau:1932,Yakovlev:2012rd},
their structure, dynamics, and composition are much richer and far more interesting. Structurally, neutron
stars contain both a solid crust and a liquid core. The non-uniform crust sits above a uniform liquid core
that largely conforms to the perception of a neutron star as a giant assembly of closed packed  neutrons,
protons, and leptons. Note that protons and leptons are fundamental stellar constituents that are required
to maintain chemical equilibrium and enforce charge neutrality.  However, the non-uniform crust deviates
drastically from such a naive picture, as it is populated by fascinating and novel states of matter. Yet, the
one kilometer stellar crust plays a rather modest role in the structure of the star. Indeed, the liquid core
accounts for practically all the mass and about 90\% of the size of a neutron star. Yet, we suggest here
that the second tidal Love number\,\cite{Binnington:2009bb,Damour:2012yf}, a quantity of great relevance
to the \emph{tidal polarizability (or deformability)} of neutron stars, is sensitive to the poorly known crustal
equation of state. Our study is motivated by the first direct detection of the binary neutron star merger
GW170817\,\cite{Abbott:PRL2017} that is providing fundamental new insights into the nature of dense
matter, primarily through constraints on the tidal polarizability. Although GW170817 has been instrumental
in elucidating the site of the $r$-process and confirming the association between binary neutron star
mergers and short-gamma ray bursts, much activity has also been devoted to explore the impact of
GW170817 on the equation of state (EOS). In particular, constraints on the tidal polarizability from
the gravitational wave profile have already been used to impose limits on stellar radii and ultimately
on the EOS\,\cite{Bauswein:2017vtn,Fattoyev:2017jql,Annala:2017llu,Abbott:2018exr,Most:2018hfd,
Tews:2018chv,Malik:2018zcf,Tsang:2018kqj,Radice:2018ozg}.

To place our work in the proper context, we now present a brief
description of the main features of the stellar crust. At the
typical densities of the \emph{outer crust}, ranging from about
$10^{4}{\rm g/cm^{3}}$ to $4\!\times\!10^{11}{\rm
g/cm^{3}}$\,\cite{Baym:1971pw}, the electrons have been pressure
ionized and move freely as a relativistic free Fermi gas. However,
as compared with the typical densities found at the center of atomic
nuclei, of the order of nuclear matter at saturation
$\rhoz\!=\!0.15\,{\rm fm}^{-3}\!\simeq\!2.5\!\times\!10^{14}{\rm
g/cm^{3}}$, these crustal densities are considerably smaller,
resulting in an average inter-nucleon separation much larger than
the range of the nucleon-nucleon interaction. Thus, it becomes
energetically favorable for nucleons to cluster into finite nuclei
which arrange themselves in a crystalline lattice. Having the lowest
mass per nucleon in the entire nuclear chart, ${}^{56}$Fe is the
most favorable nucleus at the lowest densities found in the outer
crust. However, as the density increases and with it the electronic
contribution to the energy, ${}^{56}$Fe ceases to be the nucleus of
choice. Indeed, it becomes energetically advantageous for electrons
to capture onto protons and for the excess energy to be carried away
by neutrinos. As such, the system evolves into a Coulomb crystal of
progressively more exotic neutron-rich nuclei, ranging from
${}^{62}$Ni up to (likely) ${}^{118}$Kr---a ``drip nucleus" with 32
more neutrons than its heaviest stable isotope
${}^{86}$Kr\,\cite{RocaMaza:2008ja}. The equation of state of the
outer crust is fairly simple to understand as it is dominated by the
electronic contribution with a small correction from the crystalline
lattice. The major uncertainty in the EOS emerges from the absence
of experimentally determined masses for the most exotic nuclei
populating the crystal lattice which, in turn, set the electron
fraction in the crust.

Below the outer crust sits the \emph{inner crust} spanning a region
from the neutron-drip density of $4\!\times\!10^{11}{\rm g/cm^{3}}$
up to the transition density to the uniform liquid core. At the top
layers of the inner crust nucleons continue to cluster into a
Coulomb crystal of neutron-rich nuclei embedded in a uniform
electron gas. Now, however, the crystal is also in equilibrium with
a dilute, likely superfluid, neutron vapor.  As the density
increases, the spherical nuclei that form the crystal lattice start
to deform in an effort to reduce the Coulomb repulsion, resulting in
the formation of rich and complex structures collectively referred
to as \emph{nuclear pasta}\,\cite{Ravenhall:1983uh,Hashimoto:1984}.
Moreover, due to the preponderance of quasi-degenerate low-energy
states---a hallmark of ``Coulomb frustration''---these systems
display an interesting yet subtle low-energy dynamics that has been
captured using either semi-classical simulations\,\cite{Horowitz:2004yf,
Horowitz:2004pv,Horowitz:2005zb,Watanabe:2003xu,Watanabe:2004tr,
Watanabe:2009vi,Schneider:2013dwa,Horowitz:2014xca,Caplan:2014gaa}
or quantum-mechanical mean-field approaches\,\cite{Bulgac:2001,
Magierski:2001ud,Chamel:2004in,Newton:2009zz,Schuetrumpf:2015nza,
Fattoyev:2017zhb}. Yet despite the undeniable progress in understanding 
the nuclear-pasta phase, we are unaware of any theoretical framework 
that simultaneously incorporates quantum-mechanical effects and 
dynamical correlations beyond the mean-field level. As a result, a 
reliable equation of state for the inner crust is still missing. In the past, 
we have adopted a simple polytropic interpolation formula\,\cite{Link:1999ca} 
to estimate the equation of state in the inner crust\,\cite{Carriere:2002bx}. 
The central goal of this work is to examine the sensitivity of the tidal 
Love number to the unknown EOS in the inner stellar crust.

We have organized the paper as follows. In Sec.\,\ref{Formalism} we review the essential details
required to compute the tidal polarizability and its expected sensitivity to the choice of equation
of state. In particular, special attention is paid to the component of the equation of state used to
model the inner crust. In Sec.\,\ref{Results} we provide predictions for the second Love number
and examine its sensitivity to the various choices for the equation of state. In particular, we show
that  although the crustal component of the EOS plays a minor role in the determination of both
the radius and the mass, it has a significant impact on the Love number. Finally, we summarize
and conclude in Sec.\,\ref{Conclusions}.

\section{Formalism}
\label{Formalism}

The formalism is divided into two main components, one dedicated to the description of the
underlying equation of state and the other one to the calculation of the second tidal Love
number from which the tidal polarizability can be readily computed.

\subsection{Equation of State}
\label{EOS}

\subsubsection{Liquid Core}
\label{liquidcore}

Our starting point is a particular class of relativistic effective field theories containing
as elementary constituents an isodoublet nucleon field ($\psi$) interacting via the
exchange of two isoscalar mesons---the scalar sigma ($\phi$) and the vector omega
($V^{\mu}$)---one isovector meson, the rho (${\bf b}^{\mu}$), and the photon ($A^{\mu}$).
Besides conventional Yukawa couplings\,\cite{Walecka:1974qa}, the model is supplemented
by nonlinear self\,\cite{Boguta:1977xi,Serot:1984ey,Mueller:1996pm} and
mixed\,\cite{Horowitz:2000xj,Todd-Rutel:2005fa,Chen:2014sca} interactions
between the mesons that are critical to improve the quantitative standing of the model.
The interacting Lagrangian density describing such class of effective theories is given
by\,\cite{Chen:2014sca}:
\begin{eqnarray}
{\mathscr L}_{\rm int} &=&
\bar\psi \left[g_{\rm s}\phi   \!-\!
         \left(g_{\rm v}V_\mu  \!+\!
    \frac{g_{\rho}}{2}{\mbox{\boldmath $\tau$}}\cdot{\bf b}_{\mu}
                               \!+\!
    \frac{e}{2}(1\!+\!\tau_{3})A_{\mu}\right)\gamma^{\mu}
         \right]\psi \nonumber \\
                   &-&
    \frac{\kappa}{3!} (g_{\rm s}\phi)^3 \!-\!
    \frac{\lambda}{4!}(g_{\rm s}\phi)^4 \!+\!
    \frac{\zeta}{4!}   g_{\rm v}^4(V_{\mu}V^\mu)^2 +
   \Lambda_{\rm v}\Big(g_{\rho}^{2}\,{\bf b}_{\mu}\cdot{\bf b}^{\mu}\Big)
                           \Big(g_{\rm v}^{2}V_{\nu}V^{\nu}\Big)\;.
 \label{LDensity}
\end{eqnarray}
The non-linear scalar couplings ($\kappa$ and $\lambda$) pioneered by Boguta and
Bodmer are responsible for softening the equation of state of symmetric nuclear matter
near saturation density\,\cite{Boguta:1977xi}. This softening reduces the compressibility
coefficient of symmetric nuclear matter and is instrumental in bringing theoretical
calculations of the isoscalar monopole response into agreement with experiment. The
omega-meson self-coupling $\zeta$ also softens the equation of state of symmetric
nuclear matter, but at much higher densities. By tuning this single parameter, one can
generate neutron stars with maximum masses that vary by more than one solar
mass while retaining agreement with laboratory observables\,\cite{Mueller:1996pm}.
Finally, the nonlinear mixed coupling $\Lambda_{\rm v}$ is highly sensitive to the
density dependence of symmetry energy and in particular to its slope at saturation
density---a quantity with a strong impact on the structure and dynamics of both exotic
neutron-rich nuclei and neutron stars\,\cite{Horowitz:2000xj,Horowitz:2001ya,
Carriere:2002bx,Horowitz:2004yf,Chen:2014sca,Chen:2014mza}.

In the spirit of density functional theory, the parameters of the model are directly fitted to
laboratory observables, such as binding energies and charge radii of magic and semi-magic
nuclei. Whereas most of the parameters are well calibrated by such fitting protocol, two
of them---$\Lambda_{\rm v}$ and $\zeta$---are sensitive to physics that has not yet been
efficiently probed in either laboratory experiments or astrophysical observations. This lack
of sensitivity could be remedied by probing isovector observables with a large neutron
excess, such as the thickness of the neutron skin of neutron-rich nuclei, or observables
sensitive to the high density behavior of the EOS, such as the maximum neutron-star
mass\,\cite{Fattoyev:2011ns,Fattoyev:2012rm,Chen:2014sca}. One could also constrain
these poorly determined parameters by exploring their impact on novel neutron-star
observables, such as the tidal polarizability. Yet, it is important to underscore that the
above model is used exclusively to compute the equation of state of the uniform liquid core;
that is, the phase consisting of a uniform, charged-neutral system composed of neutrons,
protons, electrons, and muons in chemical equilibrium.

The equation of state of cold neutron-rich matter describes the connection between the energy
density, the pressure, and the baryon density of the system. At zero temperature, the energy
density and pressure (both intensive quantities) are functions exclusively of the conserved
baryon density $\rho\!=\rho_{n}\!+\!\rho_{p}$ and the neutron-proton asymmetry
$\alpha\!\equiv\!(\rho_{n}\!-\!\rho_{p})/(\rho_{n}\!+\!\rho_{p})$. To identify a few of the critical
properties of the EOS it is customary to expand the energy per nucleon in even powers of
$\alpha$:
\begin{equation}
  \frac{E}{A}(\rho,\alpha) -\!M \equiv {\cal E}(\rho,\alpha)
                          = {\cal E}_{\rm SNM}(\rho)
                          + \alpha^{2}{\cal S}(\rho)
                          + {\cal O}(\alpha^{4}) \,,
 \label{EOS}
\end {equation}
where ${\cal E}_{\rm SNM}(\rho)\!=\!{\cal E}(\rho,\alpha\!\equiv\!0)$ is the energy per nucleon
of symmetric nuclear matter (SNM) and the symmetry energy ${\cal S}(\rho)$ represents to a
good approximation the energy cost of converting symmetric nuclear matter into pure neutron
matter. Also customary is to encode the behavior of both symmetric nuclear matter and the
symmetry energy in the vicinity of saturation density $\rhoz$ in terms of a few bulk parameters.
Introducing $x\!=\!(\rho\!-\!\rhoz)\!/3\rhoz$ to quantify the deviations of the EOS from its value
at saturation density we obtain\,\cite{Piekarewicz:2008nh},
\begin{subequations}
\begin{align}
 & {\cal E}_{\rm SNM}(\rho) = \epsz + \frac{1}{2}Kx^{2}+\ldots ,\label{EandSa}\\
 & {\cal S}(\rho) = J + Lx + \frac{1}{2}K_{\rm sym}x^{2}+\ldots   \label{EandSb}
\end{align}
\label{EandS}
\end{subequations}
Given that the saturation of symmetric nuclear matter is one of the hallmarks of the nuclear
dynamics, accurately-calibrated models that are informed by the binding energy and charge
radius of a variety of nuclei predict the saturation point at a density of $\rhoz\simeq\!0.15\,{\rm fm}^{-3}$
and an energy per particle of $\epsz\!\simeq\!-16\,{\rm MeV}$. As a result, the pressure of
symmetric nuclear matter vanishes at saturation, so the small density fluctuations around
the saturation point are controlled by the incompressibility coefficient $K$.

However, unlike symmetric nuclear matter, pure neutron matter does not saturate. As a result,
the pressure of pure neutron matter at saturation ($P_{0}$) does not vanish. This suggests
that the behavior of the symmetry energy around saturation density is largely contained in three,
rather than two, bulk parameters: its value $J$, its slope $L$, and its curvature $K_{\rm sym}$
at saturation density. Note that in the so-called parabolic approximation\,\cite{Piekarewicz:2008nh}
the slope of the symmetry energy $L$ is directly proportional to the pressure of pure neutron matter.
That is,
\begin{equation}
  P_{0} = \frac{1}{3}\rhoz L\;.
 \label{PZero}
\end {equation}

The Lagrangian density introduced in Eq.\,(\ref{LDensity}) will be used to compute both the
equation of state of the uniform liquid core as well as properties associated with the transition
from the liquid core to the solid crust. The development of a non-uniform crust is largely due
to the short-range nature of the strong interaction. At densities at which the inter-nucleon
separation in the uniform system becomes larger than the range of the nucleon-nucleon
interaction, it becomes energetically advantageous for nucleons to cluster into finite nuclei
in order to benefit from the nuclear attraction. To describe the crust-to-core transition one
determines the density at which the uniform liquid becomes unstable against small amplitude
density oscillations. The instability may be determined dynamically by implementing the relativistic
random-phase-approximation (RPA) described in Ref.\,\cite{Carriere:2002bx}. The outcome
of such RPA analysis is the transition density together with the associated pressure and energy
density at which the instability develops. In particular, the transition density is strongly sensitive
to the density dependence of the symmetry energy, particularly to its slope at saturation density
$L$\,\cite{Horowitz:2000xj}.

\subsubsection{Outer Crust}
\label{outercrust}

Before discussing the equation of state in the inner crust we address the simpler dynamics
of the \emph{outer} crust---the low-density region in which all neutrons are bound to finite
nuclei. The dynamics of the outer crust is encapsulated in a relatively simple expression for
the total Gibbs free energy per nucleon, which at zero temperature equals the total chemical
potential of the system. That is\,\cite{Haensel:1989,Haensel:1993zw,Ruester:2005fm,
RocaMaza:2008ja,RocaMaza:2011pk},
 \begin{equation}
   \mu(Z,A; P) = \frac{M(Z,A)}{A} + \frac{Z}{A}\mu_{e} -
   \frac{4}{3}C_{l}\frac{Z^2}{A^{4/3}}\,p_{{}_{\rm F}},
   \label{ChemPot}
\end{equation}
where $Z$ and $A$ are the atomic and mass number of the ``optimal" finite nucleus (to be
defined shortly) and P is the pressure of the system, which is dominated by the degenerate
electrons. To use the above equation it is necessary to provide a connection between the
pressure and the baryon density of the system $\rho$, or its proxy the Fermi momentum
$p_{{}_{\rm F}}$ that is defined as follows:
 \begin{equation}
   p_{{}_{\rm F}} = \left(3\pi^2\rho\right)^{1/3}.
   \label{FermiMom}
\end{equation}
The connection between the pressure and the baryon density is encoded in the equation
of state of a body-centered-cubic lattice of neutron-rich nuclei embedded in a uniform
electron gas. That is\,\cite{Baym:1971pw,RocaMaza:2008ja},
 \begin{equation}
   P(\rho) =
   \frac{m_{e}^{4}}{3\pi^{2}}
   \left(x_{{}_{\rm F}}^{3}y_{{}_{\rm F}} \!-\!
   \frac{3}{8}\left[x_{{}_{\rm F}}y_{{}_{\rm F}}
   \Big(x_{{}_{\rm F}}^{2}\!+\!y_{{}_{\rm F}}^{2}\Big)
   \!-\! \ln(x_{{}_{\rm F}}\!+\!y_{{}_{\rm F}}) \right]\right)
   \!-\! \frac{\rho}{3}C_{l}\frac{Z^2}{A^{4/3}}\,p_{{}_{\rm F}}.
 \label{Pressure}
\end{equation}
Here $x_{{}_{\rm F}}\!=\!p_{{}_{\rm F}}^{e}/m_{e}$ and
$y_{{}_{\rm F}}\!=\!(1\!+\!x_{{}_{\rm F}}^{2})^{1/2}$ are scaled electronic Fermi momentum
and Fermi energy expressed in terms of the electronic Fermi momentum
$p_{{}_{\rm F}}^{e}\!=\!(Z/A)^{1/3}p_{{}_{\rm F}}$ and $C_{l}\!=\!3.40665\!\times\!10^{-3}$
is the dimensionless lattice constant\,\cite{RocaMaza:2008ja}. The first term in
Eq.\,(\ref{ChemPot}) is independent of the pressure and  represents the entire nuclear
contribution to the chemical potential. It depends exclusively on the mass per nucleon of
the optimal nucleus populating the lattice. The second term represents the contribution from
a relativistic free Fermi gas of electrons. As such, it is strongly density dependent and
represents the major contribution to the pressure.  Finally, the last ``screening" term provides
the relatively modest lattice contribution to the pressure.

The above discussion suggests that the only unknown in the determination of  the crustal composition
is the optimal nucleus $(Z,A)_{\rm opt}$ populating the crystal lattice. To determine the optimal nucleus
we implement the following procedure for each value of the pressure $P$ and for every chosen nuclear
mass model, which often consists of thousands of nuclei. Given an individual nucleus $(Z,A)$ in the mass
table, one computes the chemical potential $\mu(Z,A; P)$, which requires the determination of the baryon
density by inverting Eq.\,(\ref{Pressure}). The optimal nucleus is defined as the one that \emph{minimizes}
the overall chemical potential. For low densities, the electronic contribution is modest so the optimal nucleus
is found to be ${}^{56}$Fe, the nucleus with the lowest mass per nucleon in the entire nuclear chart. As the
electronic contribution increases, it becomes energetically favorable for electrons to capture into protons,
thereby reducing the electron fraction albeit at the expense of increasing the neutron-proton asymmetry.
Ultimately, the crustal composition emerges from a subtle competition between the electronic contribution
that favors $Z/A\!=\!0$, and the nuclear symmetry energy which favors nearly symmetric nuclei. This
procedure continues until the chemical potential becomes equal to the mass of the neutron. Beyond this
pressure the drip line is reached as the optimal nucleus is unable to hold any more neutrons. This defines
the transition region from the outer crust to the inner crust. Note that the composition of the outer crust
determined in this manner requires knowledge of the masses of a few exotic nuclei that are presently
beyond experimental reach. To determine such masses one must rely on uncontrolled theoretical
extrapolations. To mitigate this problem machine learning is becoming an important tool to estimate,
with proper theoretical errors, the unknown masses of a variety of exotic nuclei\,\cite{Bayram:2013hi,
Utama:2015hva,Utama:2017wqe,Utama:2017ytc,Neufcourt:2018syo}. Note that most of these nuclei
are unstable under normal laboratory conditions, yet become stable in the outer crust because of the
presence of the neutralizing electron background.

\subsubsection{Inner Crust}
\label{innercrust}

Based on the previous discussion, two well determined boundaries are relevant to the calculation of
the structural properties of a neutron star: (a) the transition from the outer to the inner crust defined
as the region in which the chemical potential equals the neutron mass and (b) the transition from the
inner crust to the liquid core identified as the density at which the uniform ground state become unstable
to density fluctuations. How then does one connect these two regions? The answer to this question is
highly complex. Length scales that were well separated in both the crystalline phase, where the long-range
Coulomb interaction dominates, and in the uniform phase, where the short-range nuclear interaction
prevails, are now comparable. This gives rise to a universal phenomenon known as \emph{Coulomb
frustration}. At densities of relevance to the inner crust, Coulomb frustration drives the formation of the
nuclear-pasta phase, characterized by the emergence of complex topological structures dubbed
``nuclear pasta''\,\cite{Ravenhall:1983uh,Hashimoto:1984}; for some recent reviews on the captivating
structure and dynamics of the neutron-star crust see Refs.\,\cite{Chamel:2008ca,Bertulani:2012} and
references contained therein.

Although dynamically fascinating, the precise role of the nuclear pasta in the determination of the equation
of state is still unknown. In the past, we have followed the suggestion given in Ref.\,\cite{Link:1999ca} and
adopted a simple polytropic interpolation between the outer crust and the liquid core\,\cite{Carriere:2002bx}.
That is,
\begin{equation}
 P(\Edens)=A+B\Edens^{\,4/3},
\label{Polytrope}
\end{equation}
where $\Edens$ is the energy density of the system, and the two constants $A$ and $B$ are chosen so
that the pressure is continuous at both (a) the boundary between the inner and the outer crusts and (b) the
boundary between the inner crust and the liquid core. However, unlike the Tolman-Oppenheimer-Volkoff
equations that depend exclusively on the equation of state $P\!=\!P(\Edens)$, the corresponding differential
equation required to compute the tidal polarizability is also sensitive to its derivative, {\sl i.e.,} to the speed
of sound; see Sec.\,\ref{tidalpolarizability}.  Hence, as a simple test of the sensitivity of the tidal polarizability
to the choice of EOS in the inner crust, we adopt a cubic interpolation formula so that both the pressure and
its first derivative are continuous at both interphases. Note, however, that every change in the composition
of the outer crust is accompanied by a discontinuity in the speed of sound\,\cite{Han:2018mtj}.

In summary, the following prescription has been adopted for the equation of state. First, the equation of state
for the outer crust is defined at minimum values of $P_{\rm min}\!=\!6.08\times 10^{-15}$\,MeV/fm$^3$ and
$\Edens_{\rm min}\!=\!5.86\times 10^{-9}$\,MeV/fm$^3$ for the pressure and energy density, respectively, as
prescribed in the seminal work of Baym, Pethick, and Sutherland\,(BPS)\,\cite{Baym:1971pw}. The corresponding
maximum values for the pressure and energy density are determined by demanding that the chemical potential
introduced in Eq.(\ref{ChemPot}) be equal to the bare neutron mass. Note, however, that we modify slightly the
BPS equation of state by using the improved Duflo-Zuker mass formula\,{\cite{Duflo:1995} to compute the
composition of the outer crust. Second, given the complexity of the inner crust, we adopted an equation of state
that interpolates between the corresponding EOSs in the outer crust and in the liquid core, with the inner
crust-core boundary determined from the RPA analysis outlined above. Finally, beyond this transition region,
the equation of state in the uniform liquid core is derived from the Lagrangian density given in Eq.(\ref{LDensity}).
As an example, the equation of state predicted by the relativistic density functional ``FSUGarnet"\,\cite{Chen:2014mza}
is shown in Fig.\ref{Fig1}.  Although imperceptible in the log-log scale used in the figure, the EOS in the outer
crust displays various ``jumps" that are associated with changes in the nuclear composition.

\begin{figure}[ht]
\centering
 \includegraphics[width=.45\linewidth]{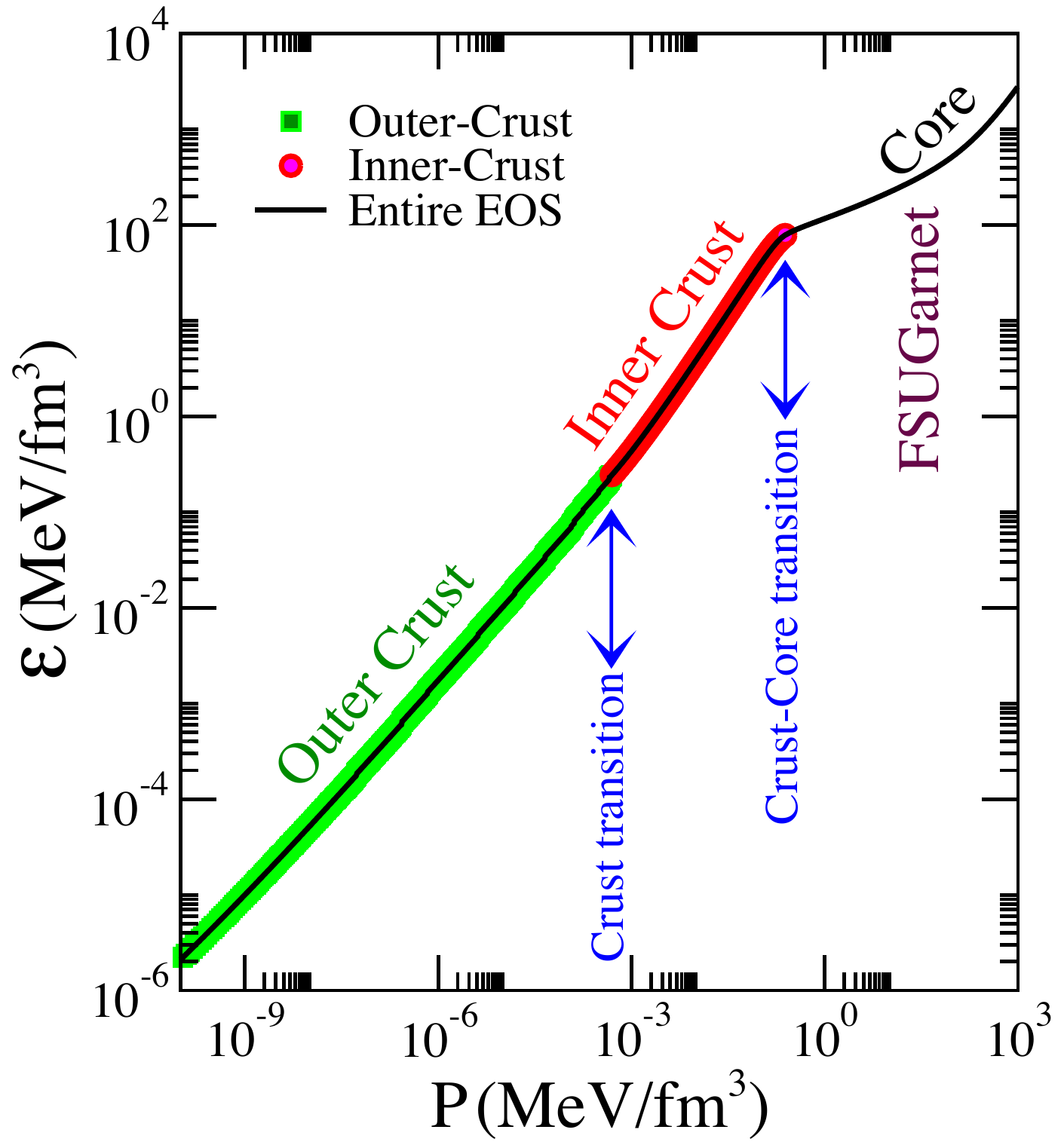}
\caption{(Color online) Neutron-star-matter equation of state as predicted by the relativistic density
functional ``FSUGarnet"\,\cite{Chen:2014mza}. The features that are sensitive to the choice of density
functional are the crust-core transition pressure and the EOS in the entire uniform liquid core.}
\label{Fig1}
\end{figure}

\subsection{Tidal Polarizability}
\label{tidalpolarizability}

As alluded in the Introduction, the first direct detection of gravitational waves from the binary neutron
star merger GW170817\,\cite{Abbott:PRL2017} is providing fundamental new insights into the nature
of dense matter. Critical properties of the equation of state are encoded in the tidal polarizability of
the neutron star, an intrinsic property that describes its tendency to develop a mass quadrupole in
response to the tidal field induced by its companion\,\cite{Damour:1991yw,Flanagan:2007ix}. As the
two neutron stars approach each other and tidal effects become important, the phase of the gravitational
wave is modified from its point-mass nature characteristic of black holes. How ``early" during the inspiral
phase do tidal effects become important is highly sensitive to the compactness of the star, and ultimately
to the underlying equation of state. For a given mass, a star with a large radius is ``fluffy" and therefore
more sensitive to tidal effects. Such a star would deform earlier in the inspiral than the correspondingly
smaller (more compact) star. One of the main findings of the LIGO-Virgo collaboration is that tidal
polarizabilities extracted from GW170817 ``disfavor equations of state that predict less compact stars".
That is, for a given stellar mass the associated radius can not be overly large\,\cite{Fattoyev:2017jql,
Annala:2017llu,Abbott:2018exr,Most:2018hfd,Tews:2018chv,Malik:2018zcf,Tsang:2018kqj}. In the linear
regime, {\sl i.e.,} in the limit of weak tidal fields, the ratio of the induced mass quadrupole to the external
tidal field defines the tidal polarizability. This is the gravitational analog to the electric polarizability of a
polar molecule in response to an external electric field.

The dimensionless tidal polarizability $\Lambda$ is defined as follows\,\cite{Abbott:PRL2017}:
\begin{equation}
 \Lambda = \frac{2}{3}k_{2}\left(\frac{c^{2}R}{GM}\right)^{5}
                 =\frac{64}{3}k_{2}\left(\frac{R}{R_{s}}\right)^{5},
 \label{Lambda}
\end{equation}
where $k_{2}$ is the second Love
number\,\cite{Binnington:2009bb,Damour:2012yf}, $M$ and $R$ are the
mass and radius of the neutron star, and $R_{s}\!\equiv\!2GM/c^{2}$
is the associated Schwarzschild radius. Clearly, $\Lambda$ is
extremely sensitive to the compactness parameter
$\xi\!\equiv\!R_{s}/R$\,\cite{Hinderer:2007mb,Hinderer:2009ca,Damour:2009vw,Postnikov:2010yn,
Fattoyev:2012uu,Steiner:2014pda}. In turn, the second Love number
$k_{2}$ depends on both $\xi$ and $\yR$---the latter a dimensionless
parameter that (as we show below) is sensitive to the entire
equation of state\,\cite{Hinderer:2007mb,Hinderer:2009ca}:
\begin{align}
 k_{2}(\xi,\yR) &= \frac{1}{20}\xi^{5}(1-\xi)^{2}\Big[(2\!-\!\yR)+(\yR\!-\!1)\xi\Big] \nonumber \\
                       &  \times \Bigg\{ \Big[(6\!-\!3\yR)+\frac{3}{2}(5\yR\!-\!8)\xi\Big]\xi
                        + \frac{1}{2}\Big[(13\!-\!11\yR)+\frac{1}{2}(3\yR\!-\!2)\xi+\frac{1}{2}(1\!+\!\yR)\xi^{2}\Big]\xi^{3}\nonumber \\
                       &\hspace{10pt}+3\Big[(2\!-\!\yR)+(\yR\!-\!1)\xi\Big](1-\xi)^{2}\ln(1-\xi)\Bigg\}^{-1}.
 \label{k2}
\end{align}
For illustrative purposes, we display the Love number  $k_{2}$ in the ``white-dwarf" limit of $\xi\!\ll\!1$\,\cite{Postnikov:2010yn}:
\begin{equation}
 k_{2}(\xi,\yR) = -\frac{1}{2}\frac{(\yR-2)}{(\yR+3)}
                       + \frac{5}{4} \frac{(\yR^{2}+2\yR-6)}{(\yR+3)^{2}}\xi
                       - \frac{5}{56} \frac{(11\yR^{3}+66\yR^{2}+52\yR-204)}{(\yR+3)^{3}}\xi^{2}
                      + {\cal O}(\xi^{3}).
 \label{k2Newton}
\end{equation}
Expanding the Love number to two higher orders ({\sl i.e.,} up to order $\xi^{4}$) yields an approximation that remains
surprisingly accurate even near the ``black-hole" limit of $\xi\!=\!1$ ($\Lambda\!=\!0$). For neutron stars having typical
values of $\xi\!\lesssim\!1/2$, such ${\cal O}(\xi^{4})$ expansion is practically exact.

We now proceed to describe a few details involved in the computation of $\yR$. More details are reserved to the
Appendix, but for an even more extensive discussion see Refs.\,\cite{Hinderer:2007mb,Hinderer:2009ca,Postnikov:2010yn}
and references contained therein. As already mentioned, an external tidal field induces a mass quadrupole in the star.
The external tidal field plus the induced stellar quadrupole combine to produce a non-spherical component to the
gravitational potential that in the limit of axial symmetry is proportional to the spherical harmonic $Y_{20}(\theta,\varphi)$.
In turn, the ``coefficient" of $Y_{20}(\theta,\varphi)$, commonly referred to as $H(r)$, is a spherically symmetric
function that encodes the dynamical changes to the gravitational potential and satisfies a linear, homogeneous,
second order differential equation\,\cite{Hinderer:2007mb}. Once the differential equation is solved, the value of $\yR$
is obtained from the logarithmic derivative of $H(r)$ evaluated at the surface of the star:
$\yR\!=\big(rH'(r)/H(r)\big)_{r=R}$. However, since all that is needed to compute the second Love number
is the logarithmic derivative of $H(r)$, it is more efficient to solve directly for $y(r)$ which, in turn, satisfies the
following non-linear, first order differential equation\,\cite{Postnikov:2010yn,Fattoyev:2012uu}:
\begin{equation}
 r\frac{dy(r)}{dr} + y^{2}(r) + F(r)y(r) +r^{2}Q(r) = 0;
 \hspace{10pt} {\rm with}\; y(0)=2 \hspace{5pt}{\rm and}\hspace{5pt} \yR\!=\!y(r\!=\!R).
 \label{DiffEqY}
\end{equation}
Expressions for both $F(r)$ and $Q(r)$ are provided in the Appendix.

\section{Results}
\label{Results}

Although it is the tidal polarizability introduced in Eq.\,(\ref{Lambda}) that has a direct connection to the
phase of the gravitational wave, we find pertinent to start by examining each of its individual components,
particularly the Love number $k_{2}(\xi,\yR)$ as well as the entire behavior of the function $y(r)$ introduced
in Eq.\,(\ref{DiffEqY}).

\subsection{The impact of the stellar crust on $\mathbf{\yR}$}
\label{sub:yR}
In this section we examine the main features of the function $y(r)$ whose value at the stellar surface $\yR\!=\!y(r\!=\!R)$
is needed for the computation of  $k_{2}$. In particular, we examine the sensitivity of $\yR$ to the stellar crust---the
relatively thin 1\,km outer region of the star. Although not central to the analysis, we assume that the TOV equations
have already been solved so that mass, pressure, and energy density profiles are readily available. This implies that both
functions $F(r)$ and $Q(r)$ given in Eq.\,(\ref{DiffEqY}) are known. Note, however, that unlike the TOV equations, $Q(r)$
depends on the speed of sound across the entire star, which is discontinuous in the outer crust due to a change in
composition. Given $F(r)$ and $Q(r)$ as defined in the Appendix, one can then proceed to solve the differential equation
for $y(r)$ using one of the many available differential solvers, such as the 4th-order Runge-Kutta method.

\begin{figure}[ht]
\centering
 \includegraphics[width=.45\linewidth]{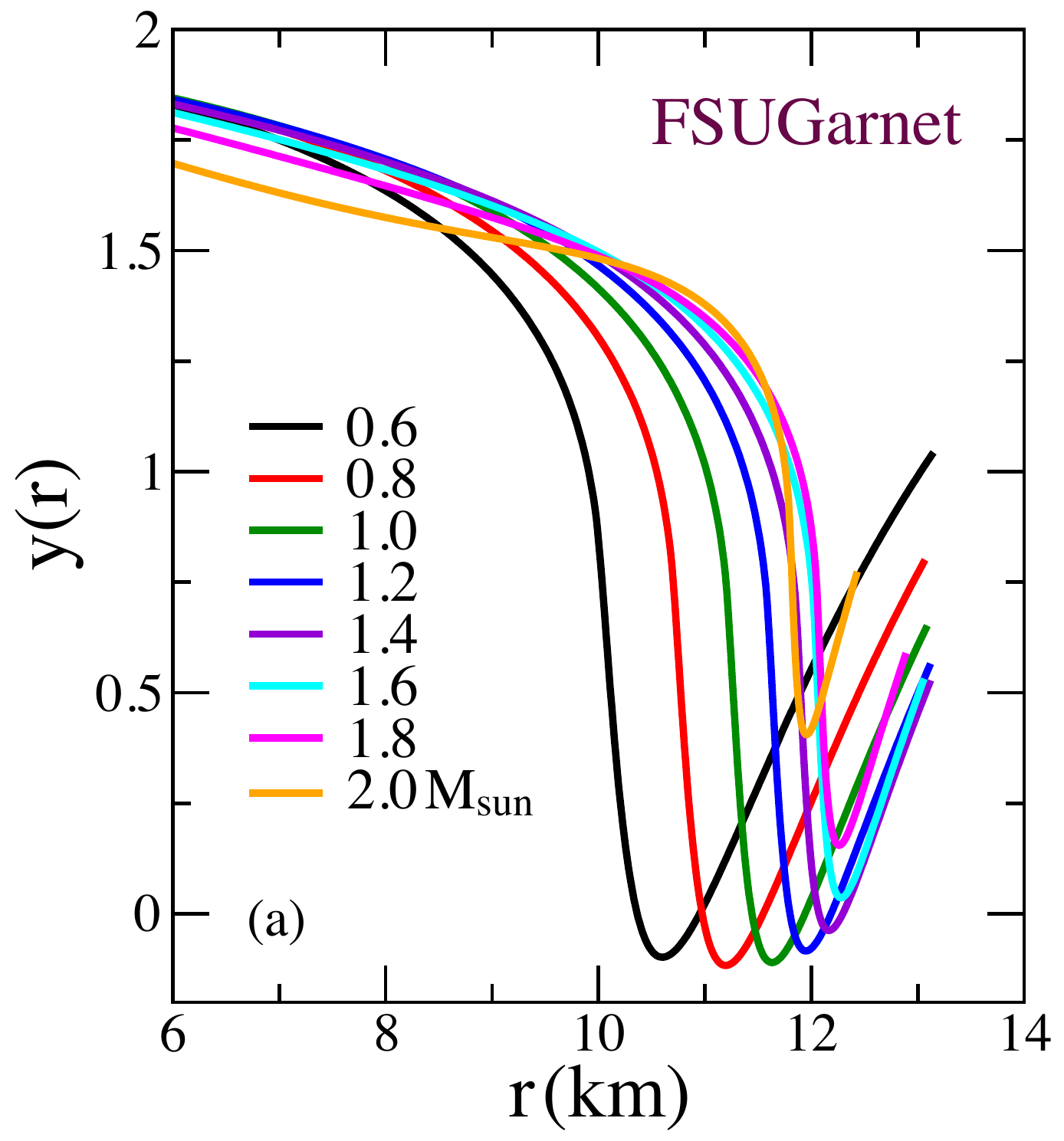}
  \hspace{10pt}
 \includegraphics[width=.45\linewidth]{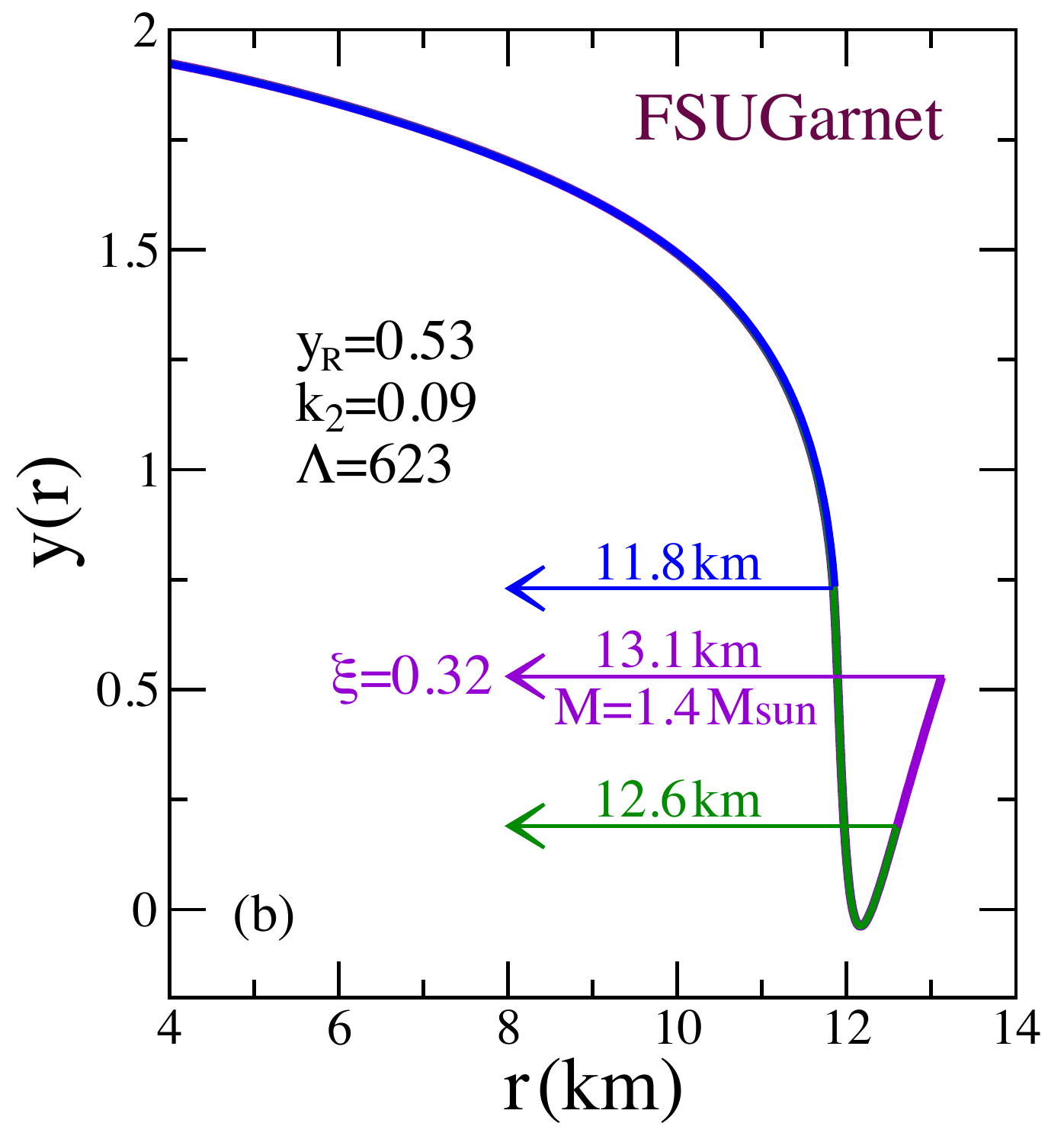}
\caption{(Color online) (a) The $y(r)$ profile as defined in Eq.\,(\ref{DiffEqY}) for a variety of neutron-star masses
as predicted by FSUGarnet\,\cite{Chen:2014mza}. The Love number is sensitive to $\yR$, namely, the value of
$y(r)$ at the surface of the star. (b) Same as in (a) but now exclusively for a $1.4\,M_{\odot}$ neutron star with
special emphasis on the contribution to $y(r)$ from various regions of the star.}
\label{Fig2}
\end{figure}

\begin{figure}[ht]
\centering
 \includegraphics[width=.50\linewidth]{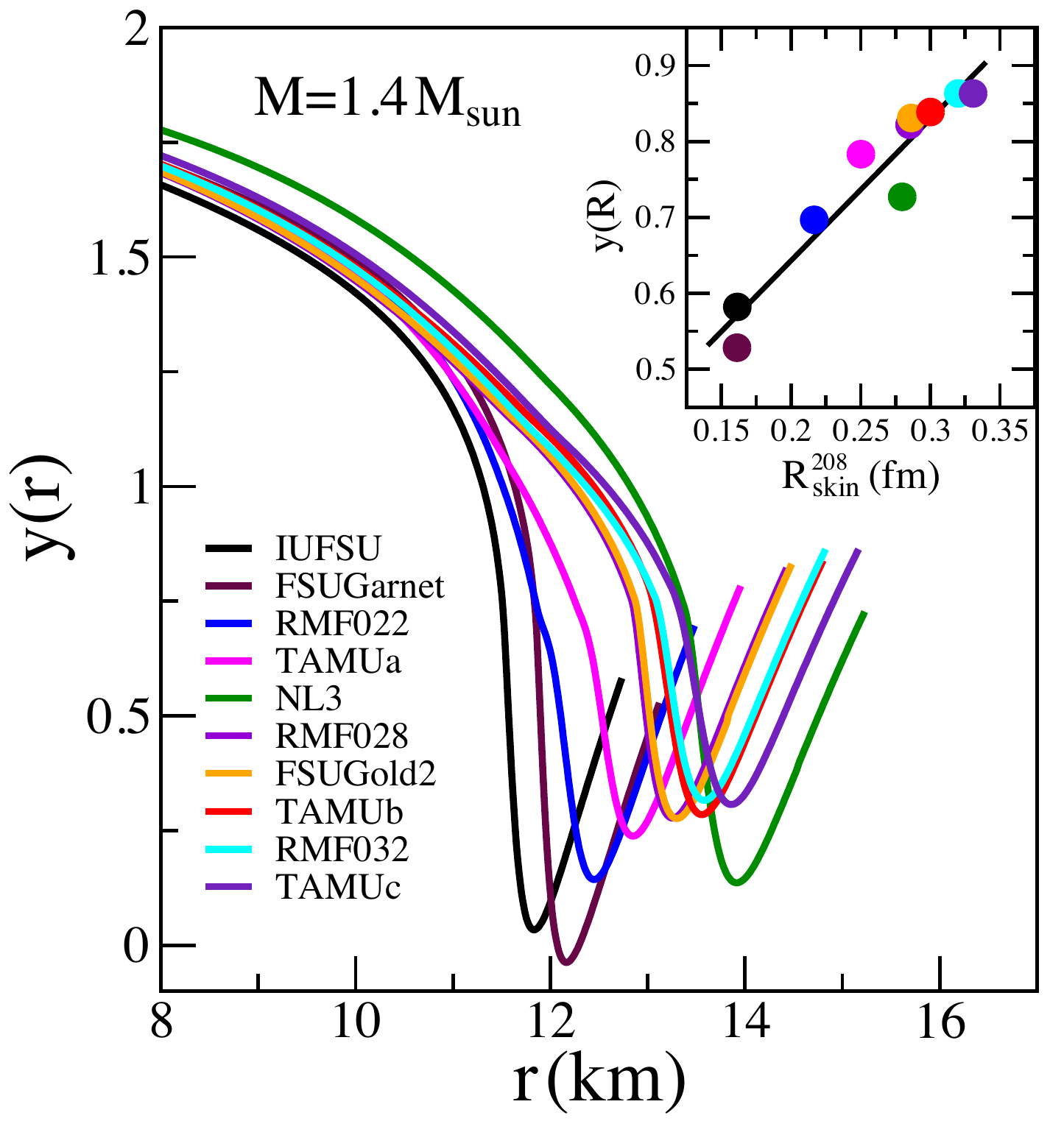}
\caption{(Color online) The $y(r)$ profile as defined in Eq.\,(\ref{DiffEqY}) for a $1.4\,M_{\odot}$ neutron star
as predicted by a collection of ten relativistic models with different choices for the density dependence of the
symmetry energy. The inset shows a correlation between $\yR\!=\!y(R)$ and the neutron skin thickness of
${}^{208}$Pb that is used as proxy for the slope of the symmetry energy $L$.}
\label{Fig3}
\end{figure}

In Fig.\,\ref{Fig2}(a) we show the entire function $y(r)$ for a
range of neutron-star masses as predicted by ``FSUGarnet", an
accurately-calibrated relativistic model that accounts for known
properties of atomic nuclei and neutron stars\,\cite{Chen:2014mza}.
The function starts at $y(0)\!=\!2$ and decreases smoothly beyond
$r\!=\!0$ until it reaches the outer stellar region where it changes
rapidly and becomes non-monotonic. This characteristic shape is
independent of the stellar mass, although changes in the function
become more pronounced with increasing mass. To better illustrate
this behavior---and to underscore the role of the stellar crust---we
isolate in Fig.\,\ref{Fig2}(b) the profile associated with a
$1.4\,M_{\odot}$ neutron star. The EOS in the uniform liquid core
dominates up to nearly 12\,km, or about 90\%, of the radius of the
star. Over this region $y(r)$ is relatively smooth and its value
drops from $y(0)\!=\!2$ at the origin to about 0.75. Beyond this
region the uniform ground state becomes unstable to cluster
formation and the solid crust develops. The $11.8\!-\!12.6\,{\rm
km}$ region comprises the entire inner crust and it is here where
the function displays its entire non-monotonic behavior. Over this
region $y(r)$ drops below zero and ``heals" to a value of 0.19 at
the interface between the inner and outer crust. The behavior of
$y(r)$ in the outer crust is fairly smooth and yields its final
value of $\yR\!=\!0.53$. Knowledge of both the compactness parameter
$\xi$ and $\yR$ is sufficient to determine the second Love number
$k_{2}\!=\!0.091$ and ultimately the tidal polarizability
$\Lambda_{1.4}\!=\!623$. We note in passing that the value of the
tidal polarizability reported here is consistent with the limit
extracted from GW170817 and reported in the original discovery
paper\,\cite{Abbott:PRL2017}. However, in a more recent analysis
that assumes that both colliding bodies are neutron stars that are
described by the same equation of state, the limit on the tidal
polarizability becomes even more stringent:
$\Lambda_{1.4}\!=\!190^{+390}_{-120}$\,\cite{Abbott:2018exr}. This
result favors soft equations of state, perhaps even softer than the
already soft FSUGarnet EOS. Moreover, it strengthens the argument
presented in Ref.\,\cite{Fattoyev:2017jql} that suggests that if the
upcoming PREX-II measurements confirms that the neutron skin
thickness of ${}^{208}$Pb is
large\,\cite{Abrahamyan:2012gp,Horowitz:2012tj}, this may be
evidence of a softening of the symmetry energy at high
densities---likely indicative of a phase transition in the stellar
interior.

We end this section by displaying in Fig.\,\ref{Fig3} $y(r)$ for a $1.4\,M_{\odot}$ neutron star as predicted by
a collection of relativistic models that, while accurate in their predictions of nuclear ground-state properties, differ
significantly in the behavior of the poorly-known density dependence of the symmetry energy\,\cite{Lalazissis:1996rd,
Lalazissis:1999,Fattoyev:2010mx,Fattoyev:2013yaa,Chen:2014sca,Chen:2014mza}. The inset in the figure shows
$\yR$ as a function of the neutron skin thickness of ${}^{208}$Pb, which is a good proxy for the density dependence
of the symmetry energy. In summary, we conclude that whereas the stellar crust plays a minor role in the determination
of the radius of the star and even less so in the determination of the mass, it greatly influences the shape of the entire
profile $y(r)$ and particularly its value at the surface $\yR$. Moreover, one should remember that the inner crust, the
region in which the behavior of $y(r)$ is no longer monotonic, is believed to harbor the nuclear-pasta phase, an exotic
state of matter with an equation of state that is presently uncertain. Thus, it is critical to quantify uncertainties in the
second Love number that originate from the various choices adopted for the EOS in the inner crust.

\subsection{The (in)sensitivity of the tidal polarizability to the EOS of the inner crust}
\label{ICandTP}

As mentioned earlier in Sec.\,\ref{innercrust}, the complexity of the inner stellar crust has hindered the 
construction of a detailed EOS and has made us rely on a cubic interpolation between the solid outer 
crust and the uniform liquid core. It was shown in Ref.\,\cite{Piekarewicz:2014lba} that the ``crustal radius"
displays a strong sensitivity to the EOS of the inner crust. We define the crustal radius as the $\sim\!1$\,km 
component of the entire stellar radius residing in the crust. In this section we examine the impact of the crustal 
EOS on $\yR$, $k_2$, and $\Lambda$. For simplicity, we choose an EOS for the inner crust of the form 
$P(\Edens)=A+B\Edens^{\gamma}$\,\cite{Link:1999ca,Carriere:2002bx}, using three different polytropic 
indices: $\gamma\!=\!1$, $4/3$, and $2$.

  \begin{table*}[th]
  \begin{tabular}{|c|c|c|c|c|c|c|}
    \hline
    $\gamma$  & $R_{\rm crust}$ & $R$ & $\xi$ & $\yR$  & $k_2$  & $\Lambda$   \\
    \hline
    \hline
    $1.0$     &  $1.402$    &  $13.246$    & $0.312$    & $0.600$ & $0.087$   & $623.7$    \\
    $4/3$     &  $0.985$    &  $12.828$    & $0.322$    & $0.342$ & $0.102$   & $623.1$    \\
    $2.0$     &  $0.755$    &  $12.598$    & $0.328$    & $0.185$ & $0.111$   & $623.2$    \\
     \hline
  \end{tabular}
 \caption{Predictions for the crustal radius, total stellar radius, compactness parameter $\xi$, 
               $\yR$, Love number $k_2$, and dimensionless tidal polarizability $\Lambda$ as a
               function of the polytropic index $\gamma$ for a $1.4 M_{\odot}$ neutron star. Both
               the crustal and total radii are given in kilometers and the EOS for the uniform liquid
               core is as predicted by the FSUGarnet model\,\cite{Chen:2014mza}.}
 \label{Table1}
 \end{table*}

In Table\,\ref{Table1} we examine the impact of the choice of polytropic index $\gamma$ on the crustal 
and total stellar radii, $\xi$, $\yR$, $k_2$, and $\Lambda$ for a $1.4 M_{\odot}$ neutron star. For all three
values of $\gamma$ the EOS for the uniform liquid core is the one predicted by the FSUGarnet 
model\,\cite{Chen:2014mza}. As is evident from these results, the EOS for the inner crust is important in 
the determination of all these quantities---except for the dimensionless tidal polarizability $\Lambda$. This 
surprising result emerges from an unexpected cancellation between the second Love number $k_{2}$ and 
the stellar compactness parameter $\xi$. Whereas $\Lambda$ depends on both $\xi$ and $k_{2}$, with $k_{2}$ 
a highly complex function of $\xi$ and $\yR$ [see Eq.(\ref{k2})], the product $k_{2}\,\xi^{-5}\!\propto\Lambda$ 
is equal for all three values of $\gamma$ to better than 0.1\%. So while the second Love number is highly 
sensitive to the crustal component of the EOS, such sensitivity disappears in the case of the tidal polarizability, 
whose behavior, as we show in the next section, is largely dictated by the EOS of the uniform liquid core. This
is not to say that the stellar crust plays no role in the determination of $\Lambda$. Quite the contrary, a 1-2\,km
contribution from the crust to the stellar radius is critical. What we have concluded is that ``reasonable" changes
to the EOS of the inner crust have no impact on the value of  tidal polarizability.


\subsection{Tidal polarizability and stellar compactness}
\label{TPandSC}

Having examined the sensitivity of the second Love number $k_{2}(\xi,\yR)$ to the underlying EOS---particularly
to the crustal component---we now proceed to explore the model dependence of the tidal polarizability $\Lambda$
defined in Eq.\,(\ref{Lambda}). We start by displaying in Fig.\,\ref{Fig4}(a) predictions for the mass-radius relation
from a representative set of RMF models that span a relatively wide range of neutron star radii. All these models are
successful in reproducing well-measured laboratory observables and, as indicated in the figure, consistent with the
$M_{\star}\!\simeq\!2\,M_{\odot}$ limit for the maximum stellar mass\,\cite{Demorest:2010bx,Antoniadis:2013pzd}.
Moreover, because of its relativistic character, they provide a Lorentz covariant extrapolation to dense matter that
ensures that the speed of sound remains below the speed of  light at all densities. For reference, the ten models
adopted in this contribution are: NL3\,\cite{Lalazissis:1996rd,Lalazissis:1999}, IU-FSU\,\cite{Fattoyev:2010mx},
TAMUC-FSU\,\cite{Fattoyev:2013yaa}, FSUGold2\,\cite{Chen:2014sca}, and FSUGarnet together with three
parametrizations denoted by RMF022, RMF028, and RMF032\,\cite{Chen:2014mza}. Note that each model is
labeled by the predicted value of the neutron skin thickness of ${}^{208}$Pb, which is a faithful proxy for the slope
of the symmetry energy. Indeed, the larger the neutron neutron skin thickness of ${}^{208}$Pb, the larger the
stellar radius\,\cite{Horowitz:2000xj,Horowitz:2001ya}. Note that in a recent analysis of the tidal polarizability
extracted from GW170817\,\cite{Abbott:PRL2017}, we used these same ten models to infer upper limits on both
the radius of a $M_{\star}\!=\!1.4\,M_{\odot}$ neutron star and the neutron skin thickness of ${}^{208}$Pb:
$R_{1.4}\!\lesssim\!13.76$\,km and $R_{\rm skin}^{208}\!\lesssim\!0.25$\,km\,\cite{Fattoyev:2017jql}, respectively.
For every mass-radius combination predicted by a given EOS, one solves for the axially-symmetric quadrupole
field at the stellar surface $\yR$, as indicated in the Appendix. From these three quantities ($R$, $M$, and $\yR$)
both the second Love number $k_{2}$ and the dimensionless tidal polarizability $\Lambda$ can be computed.
Having done so, we display in Fig.\,\ref{Fig4}(b) the tidal polarizability as a function of both the stellar mass and
stellar radius as predicted by the various equations of state. In particular, we indicate with an arrow the upper limit
of $\Lambda_{1.4}\!\le\!800$ extracted from the original discovery paper\,\cite{Abbott:PRL2017}; this upper limit
decreases even further if one adopts the revised analysis presented in Ref.\,\cite{Abbott:2018exr}. Indicated with
circles of the appropriate color is the location of a $M_{\star}\!=\!1.4\,M_{\odot}$ neutron star. Although slightly
difficult to discern in this logarithmic scale, only the four models with the softest symmetry energy---and thus
predicting the four most compact configurations---remain consistent with the $\Lambda_{1.4}\!\le\!800$ limit;
see also Fig.4 in Ref.\,\cite{Fattoyev:2017jql}.

\begin{figure}[ht]
\centering
 \includegraphics[width=.48\linewidth]{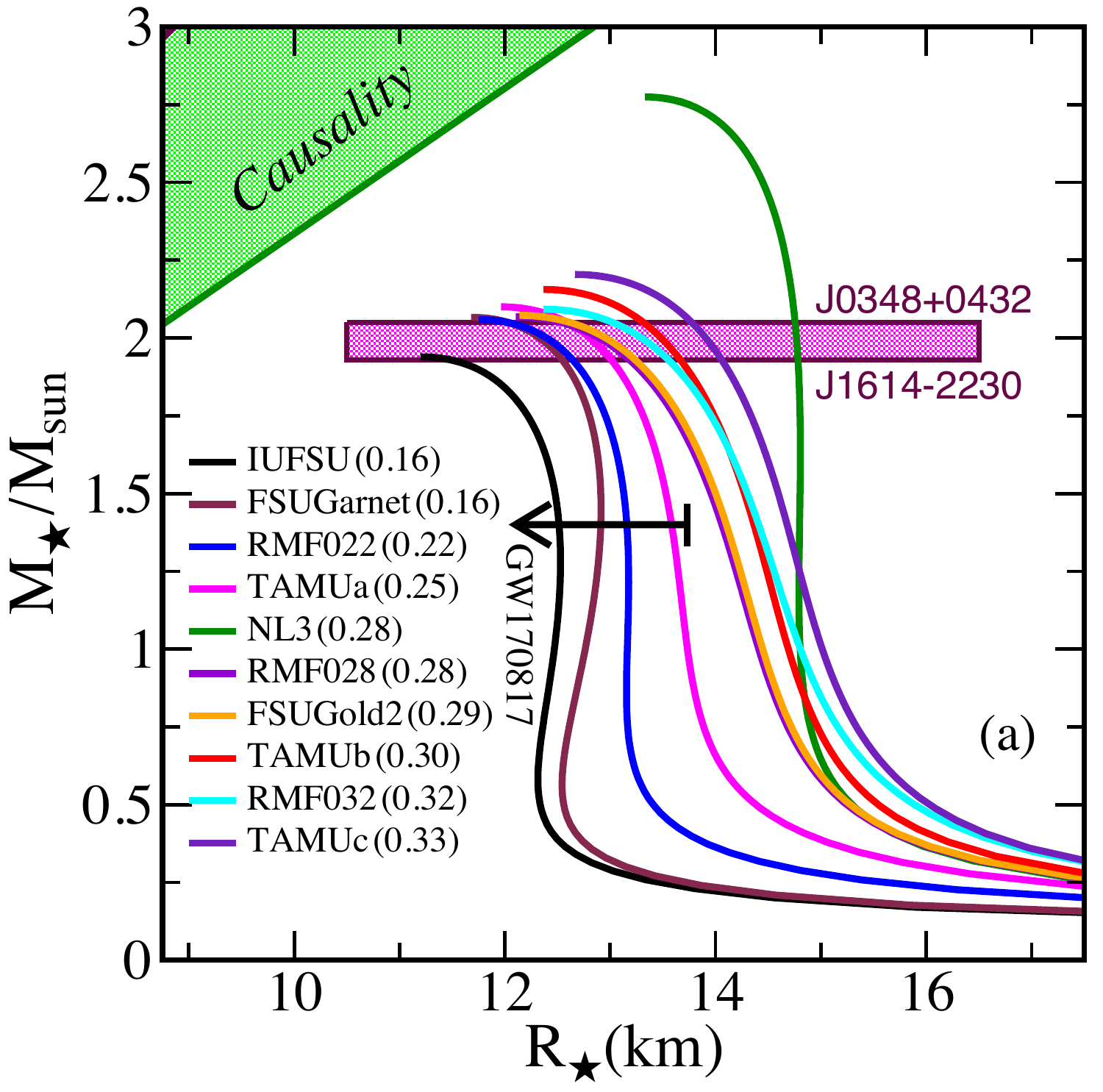}
  \hspace{5pt}
 \includegraphics[width=.49\linewidth]{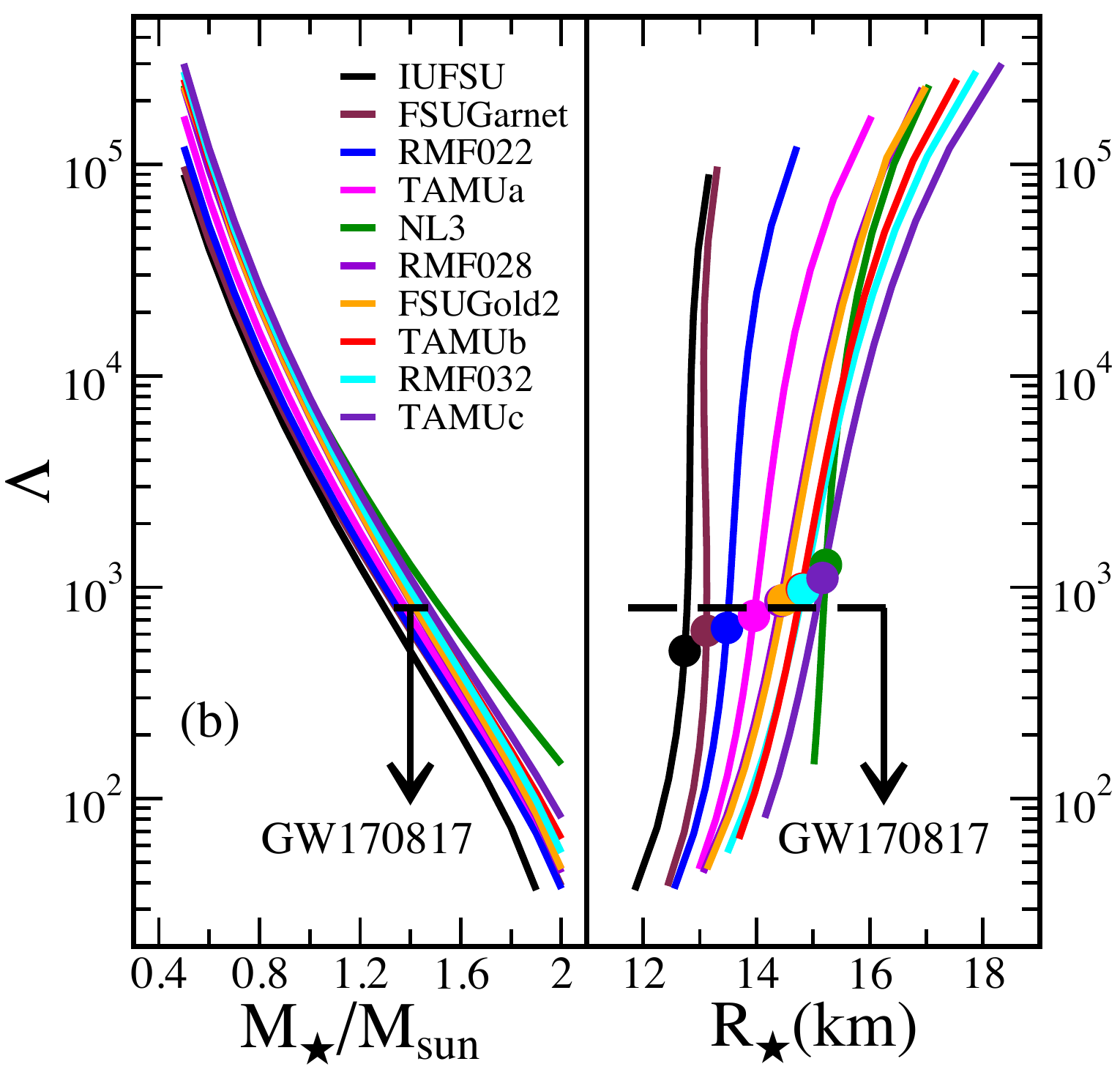}
\caption{(Color online) (a) Mass-Radius relationship predicted by the ten models of the equation of state
discussed in the text. Mass constraints from\,\cite{Demorest:2010bx,Antoniadis:2013pzd} are indicated
with a combined uncertainty bar and the radius constraint by an arrow as in Ref.\,\cite{Fattoyev:2017jql}.
The excluded causality region was adopted from Ref.\,\cite{Lattimer:2006xb}. (b) The dimensionless tidal
polarizability as a function of both the stellar mass and radius. The arrows indicate constraints inferred
from the $\Lambda\!\le\!800$ limit for a $1.4\,M_{\odot}$ neutron star\,\cite{Abbott:PRL2017}. In turn, the
colored circles indicate the corresponding predictions for the location of a $1.4\,M_{\odot}$ neutron star.}
\label{Fig4}
\end{figure}

As alluded earlier in Sec.\,\ref{tidalpolarizability}, the tidal polarizability is a very sensitive function of the
compactness parameter $\xi$. Scaling as the fifth power of $\xi$, a change in the stellar radius from
12\,km to 14\,km  leads to an increase in $\Lambda$ by more than a factor of two; see Fig.\,\ref{Fig4}(b).
However, although often used as a constraint on stellar radii, the tidal polarizability is not an independent
function of the mass and the radius. Rather, the \emph{dimensionless} tidal polarizability scales as the
dimensionless ratio of the stellar mass to the stellar radius $\xi\!\equiv\!R_{s}/R\!=\!2GM/c^2R$. There are,
of course, scaling violations to $\Lambda$ encoded in the second Love number $k_{2}$ which although
sensitive to $\xi$, also depends on the entire equation of state through its dependence on the quadrupole
field at the stellar surface $\yR$; see Eq.\,(\ref{k2}). However, for a given compactness parameter $\xi$,
the dependence of $k_{2}$ on $\yR$ is relatively mild---at least for the set of models considered in this
work. The model dependence of $k_{2}$ on the EOS is displayed in Fig.\,\ref{Fig5}(a) as a function of
the compactness parameter $\xi$. The failure of all predictions to collapse into one single curve is due
to their dependence on $\yR$. Nevertheless, the model dependence is mild, with the largest difference
between models being of the order of 25\%. Yet such mild model dependence all but disappears as one
plots the dimensionless tidal polarizability $\Lambda$ as a function of the compactness parameter $\xi$.
That in this case all ten predictions collapse into a single curve is due to the $\xi^{-5}$ scaling of $\Lambda$;
see Fig.\,\ref{Fig5}(b) where $\Lambda$ is plotted using both logarithmic and linear scales. For example,
assuming a dimensionless tidal polarizability of $\Lambda_{\star}\!=\!800$, results in theoretical predictions
from all ten models that fall within the very narrow range of $0.29\lesssim\!\xi\!\lesssim0.30$. Fig.\,\ref{Fig5}(b)
demonstrates that an accurate measurement of the dimensionless tidal polarizability fixes the stellar compactness.
This kind of ``universal" relation among various stellar observables is reminiscent of the ``I-Love-Q" relations 
explored extensively by Yagi and Yunes to project out any uncertainties in nuclear physics\,\cite{Yagi:2013bca,
Yagi:2013awa,YAGI20171}. In this context our aim is rather different. Our goal is to actually confront these nuclear 
physics uncertainties by combining laboratory experiments and astrophysical observations in an effort to determine 
the EOS. Yet, in an effort to generate the entire mass-radius relationship from which the neutron-star matter EOS 
may be uniquely determined\,\cite{Lindblom:1992}, additional stellar observables that are sensitive to a different 
combination of masses and radii must be measured. In the context of binary neutron star mergers, the ``chirp" 
mass and the ``reduced" tidal polarizability\,\cite{Flanagan:2007ix,Favata:2013rwa} offer an attractive alternative.

\begin{figure}[ht]
\centering
 \includegraphics[width=.465\linewidth]{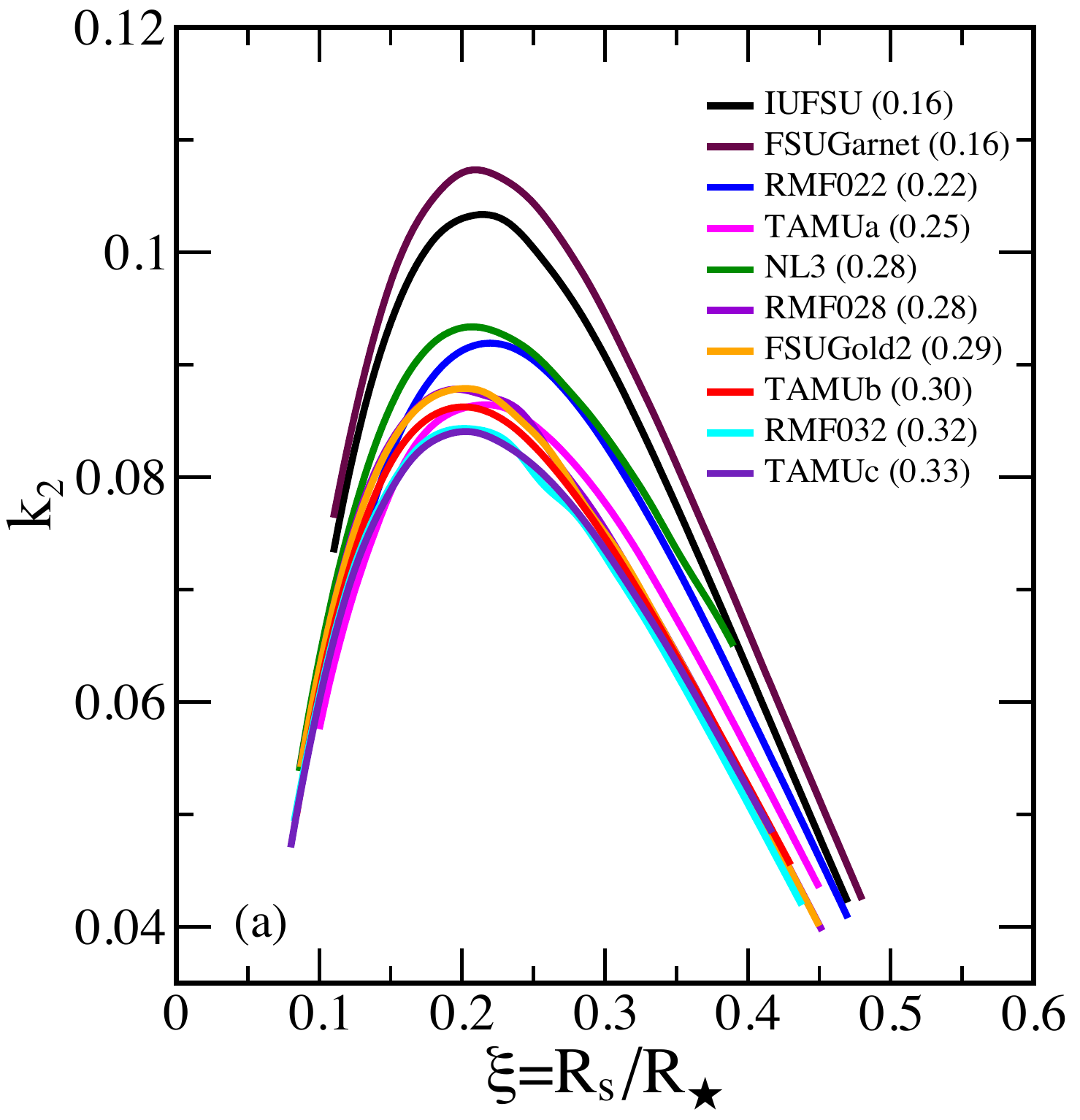}
  \hspace{10pt}
 \includegraphics[width=.475\linewidth]{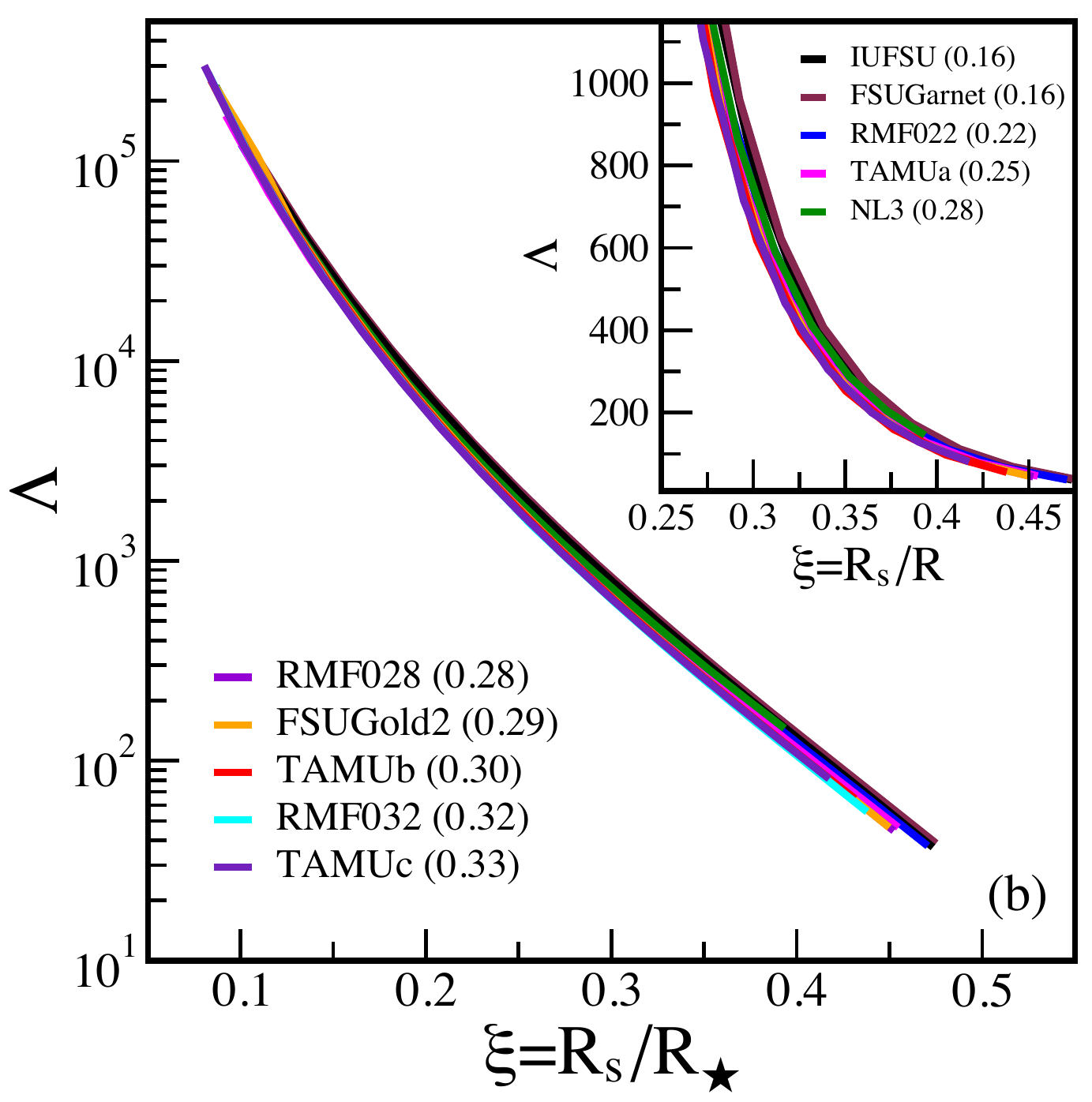}
 \caption{(Color online) (a) The second tidal Love number $k_{2}$ as a function of the compactness
  parameter $\xi$ as predicted by the ten models of the equation of state used in the text. Although
  $k_{2}$ depends on $\xi$, the failure of all predictions to collapse into a single curve is due to its
  dependence on $\yR$, which in turns depends on the entire EOS. (b) Same as (a) but now for the
  dimensionless tidal polarizability $\Lambda$. The collapse of all ten predictions into a universal
  curve is due to the strong sensitivity of $\Lambda$ to the compactness parameter $\xi$.}
\label{Fig5}
\end{figure}

\section{Conclusions}
\label{Conclusions}

The brand new era of multimessenger astronomy started on August 17, 2017 with the first
direct detection of both gravitational and electromagnetic radiation from the binary neutron
star merger GW170817. GW170817 established for the first time the association of short
gamma ray bursts with neutron star mergers and the critical role that the radioactive decay
of $r$-process elements plays in powering the kilonova light curve. Moreover, GW170817
has started to provide fundamental new insights into the nature of dense matter by adding
the tidal polarizability to the arsenal of observables that inform the equation of state. Similar
to the electromagnetic response of a polar molecule to the presence of an external electric
field, the tidal polarizability encodes the response of the neutron star to the external tidal field
produced by its companion. Given that the tidal polarizability ``hides" within a high order
coefficient in the post-Newtonian expansion of the gravitational wave form, its extraction
becomes a challenging proposition. As such, GW170817 could only established upper
limits on the tidal polarizability of a 1.4 solar mass neutron star. Nevertheless, these limits
are already stringent enough to rule out equations of state that predict relatively large stellar
radii.

Besides its sensitivity to the stellar compactness, the tidal polarizability depends on the
second tidal Love number $k_{2}$. Particularly relevant to the determination of $k_{2}$
is the value of the logarithmic derivative of the non-spherical (quadrupole) component
of the gravitational potential at the stellar surface $\yR\!\equiv\!y(r\!=\!R)$, a quantity that
emerges from solving a non-linear differential equation that is highly sensitive to the
underlying equation of state. Although the value of $\yR$ (and the compactness parameter)
is all that is needed to compute $k_{2}$, we found that the underlying function $y(r)$ displays
an interesting structure. In particular, $y(r)$ displays a non-monotic behavior that is entirely
controlled by the equation of state in the inner stellar crust. Given that the inner crust is
characterized by the emergence of complex topological structures collectively known as
nuclear pasta, the equation of state in this region is poorly known. So despite the fact
that the inner crust typically accounts for less than 10\% of the stellar radius, its impact on
the determination of $k_{2}$ was found to be significant. Whereas $y(r)$ falls
smoothly and monotonically over the uniform stellar core, its behavior in
the inner crust is neither monotonic nor smooth. Indeed, $y(r)$ attains its minimum value
(often negative) in the inner crust to then rise monotonically over the dilute regions of the
inner crust and over the entire outer crust. How will this behavior be modified, if any, with
a more realistic description of the EOS in the inner crust provides one more reason to
elucidate the fascinating and complex dynamics of the Coulomb frustrated nuclear pasta.
Nevertheless, we found that due to an unexpected cancellation between $k_{2}$ and the 
stellar compactness $\xi$, ``reasonable" modifications to the EOS of the inner stellar crust 
have no impact on the tidal polarizability.

In summary, we have examined in considerable detail the sensitivity of the second tidal Love
number to the neutron star matter equation of state. For a fixed compactness parameter,
$k_{2}$ is the only component of the tidal polarizability sensitive to the underlying EOS,
particularly to the contribution from the inner stellar crust. Given that the tidal polarizability
scales as the fifth power of the compactness parameter, $k_{2}$ provides a small correction
to the scaling relation between $\Lambda$ and the compactness parameter. Yet the
multimessenger era is in its infancy and many more detections of binary neutron star
mergers are anticipated as LIGO-Virgo prepares for its third observing run, likely to start
at the beginning of 2019. Also in 2019, PREX-II will provide significantly improved limits on
the neutron skin thickness of ${}^{208}$Pb which, in turn, will constrain the EOS in the
vicinity of saturation density. We are confident that PREX-II in conjunction with the increased
sensitivity of gravitational wave detectors will provide critical information on individual stellar
masses and tidal polarizabilities. This powerful synergy will yield a determination of the
mass-radius relation and ultimately of the underlying equation of state.

\appendix*
\section{The Love Number}
\label{Appendix}

As shown in Eq.\,(\ref{Lambda}), once the TOV equations are solved and the compactness
parameter $\xi$ has been extracted, the only remaining unknown in the computation of the
tidal polarizability is the second Love number $k_{2}(\xi,\yR)$. In this Appendix we outline the
steps necessary to compute $\yR$.

We start by invoking a Newtonian description of the static gravitational
potential in terms of two non-spherical contributions: (i) an external tidal field (perhaps
produced by the companion) plus (ii) an induced stellar contribution in response to the
external tidal field. One assumes that the external gravitational potential is slowly varying
over the dimensions of the star so that it may be expanded in a Taylor series around its
center, assumed to be the origin. The external gravitational potential at an observation
point ${\bf r}$ in the stellar neighborhood may then be written as
\begin{equation}
 \Phi_{\rm ext}({\bf r}) = \Phi_{\rm ext}(0) + {\bf r}_{i}\partial_{i}\Phi_{\rm ext}(0)
 + \frac{1}{2} {\bf r}_{i}{\bf r}_{j}\partial_{i}\partial_{j}\Phi_{\rm ext}(0) + \ldots
 = \Phi_{\rm ext}(0) + {\bf r}_{i}\partial_{i}\Phi_{\rm ext}(0)
 +\frac{1}{2}{\mathcal E}_{ij} {\bf r}_{i}{\bf r}_{j} + \ldots
  \label{PhiExt}
\end{equation}
where the external tidal field has been defined as
${\mathcal E}_{ij}\!\equiv\!\partial_{i}\partial_{j}\Phi_{\rm ext}(0)$.  Along the same lines,
the induced gravitational potential at a point ${\bf r}$ outside the star may be expanded
in a multipole series:
\begin{equation}
 \Phi_{\rm ind}({\bf r}) = -G\left(\frac{M}{r} + \frac{{\bf p}_{i}{\bf r}_{i}}{r^{3}} +
  \frac{1}{2}Q_{ij}\frac{{\bf r}_{i}{\bf r}_{j}}{r^{5}} + \ldots\right),
  \label{PhiInd}
\end{equation}
where ${\bf p}$ and $Q_{ij}$ are the dipole and quadrupole moments of the stellar distribution,
namely,
\begin{equation}
   {\bf p} = \int\!{\bf r}\rho({\bf r})d^{3}r \hspace{5pt} {\rm and} \hspace{5pt}
  Q_{ij} = \int\!\Big(3{\bf r}_{i}{\bf r}_{j}-r^{2}\delta_{ij}\Big)\rho({\bf r})d^{3}r.
  \label{PhiInd}
\end{equation}
Unlike the electric dipole moment of a charge distribution, the dipole moment of a mass distribution
can always be made to vanish by placing the origin at the center of mass. Moreover, given that the
gravitational potential is defined up to an overall constant and that the second term in
Eq.\,(\ref{PhiExt}) induces an overall translation of the center of mass, the overall gravitational potential
may be written as
\begin{equation}
 \Phi({\bf r})=\Phi_{\rm ind}({\bf r})+\Phi_{\rm ext}({\bf r}) =
  -G\left(\frac{M}{r} + \frac{1}{2}Q_{ij}\frac{{\bf r}_{i}{\bf r}_{j}}{r^{5}} + \ldots\right)
  + \frac{1}{2}{\mathcal E}_{ij} {\bf r}_{i}{\bf r}_{j} + \ldots
  \label{Phi}
\end{equation}

In the context of general relativity, the gravitational potential of a Schwarzschild star---namely, a spherical,
static, and relativistic star described by the TOV equations---is directly related to the ``tt-component" of the
metric tensor. That is,
\begin{equation}
 g_{tt}(r)\!=\!-e^{2\Phi(r)}\, \xRightarrow[\Phi\ll 1]{}    -1-2\Phi(r)+\ldots,
 \label{gttWeak}
\end{equation}
In the presence of a gravitational field with non-spherical components that are treated in the linear
regime, the Schwarzschild metric is suitably modified as follows\,\cite{Hinderer:2007mb}:
\begin{equation}
 g_{tt}(r,\theta,\phi) = -e^{2\Phi(r)}\Big(1+H(r)Y_{20}(\theta,\varphi)\Big),
  \label{gtt}
\end{equation}
where only an axially-symmetric external quadrupole field is considered and $H(r)$ satisfies a linear,
homogeneous, second-order differential equation\,\cite{Hinderer:2007mb,Hinderer:2009ca,Postnikov:2010yn}:
\begin{equation}
 \frac{d^{2}H(r)}{dr^{2}} + \left(\frac{1+F(r)}{r}\right)\frac{dH(r)}{dr} + Q(r)H(r) = 0,
 \hspace{10pt} {\rm with} \; \lim_{r\rightarrow 0} H(r) \simeq r^{2}.
 \label{Hofr}
\end{equation}
Note that the relevant bilinear product of ${\bf r}$ written in cartesians coordinates in Eq.\,(\ref{Phi}) is related
to the azimuthally independent ($l\!=\!2,m\!=\!0$) spherical harmonic:
\begin{equation}
  \big({\bf r}\otimes{\bf r}\big)_{20} = \frac{1}{\sqrt{6}}\Big(3z^{2}-r^{2}\Big) =
  \sqrt{\frac{8\pi}{15}}\hspace{2pt}r^{2}Y_{20}(\theta,\varphi).
  \label{Y20}
\end{equation}
Also note that both $F(r)$ and $Q(r)$ given in Eq.\,(\ref{Hofr}) are \emph{known} functions of the mass, pressure,
and energy density profiles assumed to have been obtained previously by solving the TOV equations and are
give by the following expressions\,\cite{Fattoyev:2012uu}:
 \begin{align}
 & F(r)  = \frac{1-4\pi Gr^{2}\Big(\Edens(r)-P(r)\Big)}
              {\displaystyle{\left(1-\frac{2GM(r)}{r}\right)}},  \\
 & Q(r)  =  \frac{4\pi}{\displaystyle{\left(1-\frac{2GM(r)}{r}\right)}}
                \left(5\Edens(r)+9P(r)+\displaystyle{\frac{\Edens(r)+P(r)}{\mathlarger{c}_{\rm s}^{2}(r)}}
                -\frac{6}{4\pi r^{2}}\right)
                - 4\left[\frac{G\Big(M(r)+4\pi r^{3}P(r)\Big)}{\displaystyle{r^{2}\left(1-\frac{2GM(r)}{r}\right)}}\right]^{2},
  \label{FandQ}
\end{align}
where $\mathlarger{c}_{\rm s}^{2}(r)\!=\!dP(r)/d\Edens(r)$ is the speed of sound at a depth $r$.

In principle, this is sufficient to compute $\yR$, that is obtained from the logarithmic derivative of $H(r)$
evaluated at the surface of the star\,\cite{Hinderer:2007mb}, namely,
\begin{equation}
 \yR = y(r\!=\!R) \equiv \left(\frac{d\ln H(r)}{d\ln r}\right)_{\!r=R} = \left(\frac{rH'(r)}{H(r)}\right)_{\!r=R}.
  \label{yR}
\end{equation}
However, given that all that is needed to compute the Love number is $\yR$, it was suggested in
Ref.\,\cite{Postnikov:2010yn} that perhaps solving directly for $y(r)$ may be more efficient than
solving for $H(r)$. That is, by introducing the following transformation,
\begin{equation}
   y(r) \equiv  \left(\frac{rH'(r)}{H(r)}\right) \iff H(r)=H(r_{0})\exp\left(\int_{r_{0}}^{r}\frac{y(r')}{r'}dr'\right),
   \label{ytoH}
\end{equation}
it is easy to show that the function $y(r)$ satisfies the following first-order---albeit non-linear---differential
equation:
\begin{equation}
 r\frac{dy(r)}{dr} + y^{2}(r) + F(r)y(r) +r^{2}Q(r) = 0, \hspace{10pt} {\rm with}\; y(0)=2.
 \label{yofr}
\end{equation}
In this way $\yR$ is directly obtained from the value of the function $y(r)$ evaluated at $r\!=\!R$.


\begin{acknowledgments}
 This material is based upon work supported by the U.S. Department
 of Energy Office of Science, Office of Nuclear Physics under Award
 Number DE-FG02-92ER40750.
\end{acknowledgments}

\bibliography{./Love2018_v2.bbl}

\end{document}